\documentclass[aps,prb,twocolumn,groupedaddress,showpacs,floatfix,
               superscriptaddress, nofootinbib]{revtex4-2}

\usepackage[colorlinks = true,
            linkcolor = red,
            urlcolor  = red,
            citecolor = blue,
            anchorcolor = blue]{hyperref}
\usepackage{epsfig}
\usepackage{amsmath,amssymb}
\usepackage{graphicx}
\usepackage[dvipsnames,usenames]{xcolor}
\usepackage[normalem]{ulem}
\usepackage{soul}  
\usepackage{glossaries}
\newacronym{HS}{HS}{Hubbard-Stratonovich}
\newacronym{ee}{$e$-$e$}{electron-electron}
\newacronym{eph}{$e$-ph}{electron-phonon}
\newacronym{DQMC}{DQMC}{determinant quantum Monte Carlo}
\newacronym{HMC}{HMC}{hybrid Monte Carlo}
\newacronym{EFA-HMC}{EFA-HMC}{exact Fourier acceleration HMC}
\newacronym{QMC}{QMC}{quantum Monte Carlo}
\newacronym{SSH}{SSH}{Su-Schrieffer-Heeger}
\newacronym{SSHH}{SSHH}{SSH-Hubbard}
\newacronym{KVBS}{KVBS}{Kekul{\'e} valence bond solid}
\newacronym{AFM}{AFM}{antiferromagnetic}
\newacronym{BOW}{BOW}{bond-order-wave}
\newacronym{CDW}{CDW}{charge-density-wave}
\newacronym{DEAC}{DEAC}{differential evolution for analytic continuation}
\newacronym{AC}{AC}{analytic continuation}
\newacronym{MaxEnt}{MaxEnt}{maximum entropy}
\newacronym{rms}{rms}{root-mean-square}
\newacronym{2D}{2D}{two-dimensional}
\usepackage{todonotes}

\usepackage{orcidlink}

\emergencystretch=\maxdimen
\begin{document}

\title{The half-filled optical Su-Schrieffer-Heeger-Hubbard model with uniaxial strain}
\author{\mbox{Jonah Huang}} 
\affiliation{Department of Physics, University of California, Davis, CA 95616, USA}
 \author{\mbox{James Neuhaus}\orcidlink{0000-0001-6904-8510}} 
 \affiliation{Department of Physics and Astronomy, The University of Tennessee, Knoxville, Tennessee 37996, USA}
\author{\mbox{Benjamin Cohen-Stead\orcidlink{0000-0002-7915-6280}}} 
\affiliation{Department of Physics and Astronomy, The University of Tennessee, Knoxville, Tennessee 37996, USA}
\affiliation{Institute for Advanced Materials and Manufacturing, The University of Tennessee, Knoxville, Tennessee 37996, USA\looseness=-1}
\author{Steven Johnston\orcidlink{0000-0002-2343-0113}} 
\affiliation{Department of Physics and Astronomy, The University of Tennessee, Knoxville, Tennessee 37996, USA}
\affiliation{Institute for Advanced Materials and Manufacturing, The University of Tennessee, Knoxville, Tennessee 37996, USA\looseness=-1}
\author{\mbox{Richard Scalettar\orcidlink{0000-0002-0521-3692}}} 
\affiliation{Department of Physics and Astronomy, University of California, Davis, California 95616, USA}

\date{\today}

\begin{abstract}
Strain offers a direct route to control electronic phases by altering bond lengths, hopping amplitudes, and lattice symmetries. These aspects make \gls*{eph} coupled systems a natural setting to study strain-induced effects. The \gls*{SSH} Hamiltonian describes the coupling between itinerant electrons and lattice degrees of freedom that modulate hopping and is thus expected to be strongly affected by strain fields. At half-filling and on a bipartite lattice, the \gls*{SSH} interactions drive dominant \gls*{BOW} correlations at low-temperature in which short bonds with high kinetic energy alternate with long bonds with low kinetic energy along four possible patterns. We model uniaxial strain through anisotropic hopping, which breaks the $90^\circ$ rotational symmetry and reduces the \gls*{BOW} phase degeneracy to two. We also include an on-site Hubbard $U$ that promotes antiferromagnetic correlations that are also (weakly) driven by the \gls*{SSH} interaction itself. We map out the half-filled phase diagram in the plane of the \gls*{eph} and Hubbard couplings, and determine the critical temperature of \gls*{BOW} formation. Using analytic continuation, we also evaluate the spectral function and demonstrate directionally dependent gap formation in the \gls*{BOW} phase.
\end{abstract}

\maketitle

\glsresetall

\section{Introduction} 

Advances in heterostructure and thin-film growth capabilities 
have established strain as a powerful means of controlling electronic and collective phases in quantum materials~
\cite{tsymbal2012multifunctional, Lotsch2015vertical, Ha2024thinfilms, Kim2023StrainEngineering}. 
Well-known examples include the formation and manipulation of a \gls*{KVBS} in strained graphene~\cite{Naumis2023Mechanical}, as
well as strong strain tuning of superconductivity in NbSe$_2$~\cite{Chen2020strain} and Sr$_2$RuO$_4$~\cite{Steppke2017StrainSr2RuO4}.
Moiré superlattices and the emerging field of twistronics provide still further examples
~\cite{Cao2018Unconventional, Mellado2025, Lu2026numerical}. 

Electronic structure methods such as density functional theory can provide enormous insight into the properties of such materials, particularly by incorporating the microscopic details of chemical composition and crystal structure~\cite{JonesGunnarsson1989, Becke2014, Jones2015, Jain2016, Maurer2019}. These approaches, however, are less suited to exploring collective phenomena arising from strong interactions; in such cases, one typically turns to a variety of model Hamiltonians~\cite{Dagotto1994, Imada1998, Kotliar2006, powell2010, Arovas2022}. 
Of these, the Hubbard model focuses on (on-site) \gls*{ee} interactions, and the ensuing Mott insulator phase as well as various magnetic, charge, and ($d$-wave) superconducting orders~\cite{georges96, Maier2005systematic, LeeWen2006, Arovas2022, Qin2022, jiang2025competition}. The Holstein~\cite{holstein59} and Su-Schrieffer-Heeger (SSH)~\cite{Barisic1970tightbinding, su79} Hamiltonians, on the other hand, describe two fundamental aspects of \gls*{eph} interactions. In the former case, the lattice displacement couples locally to the density of fermions on a site. In the latter case, the phonons modify the kinetic energy between sites by modulating the electronic hopping integrals. Both the Holstein~\cite{scalettar89, Noack1991, Vekic1992, Zhang2019, CChen2019, Feng2020, Weber2018, Cohen-Stead2019, ZXLi2019, Feng2020i, Nosarzewski2021, Xiao2021, Bradley2021, dee2020relative, Nosarzewski2021, Cohenstead2022, Bradley2023, Kvande2023, CohenStead2023Charge, Ying2024, Issa2025} and \gls*{SSH}~\cite{johnsousmonaberciu2018, XingBatrouni2021,CaiYao2021, Gotz2022, Costa2023, Ly2023, Xing2023, Casebolt2024, Costa2024Kekule, Ly2025}
models have been extensively studied with different \gls*{QMC} methods, 
with the latter including the bond-\cite{Sengupta2003Peierls}, optical~\cite{Capone1997small}, and acoustic ~\cite{Barisic1970tightbinding} variants. (We refer the reader to Ref.~\cite{Costa2023} for a detailed discussion of the differences between these models.)  

Although more challenging, \gls*{QMC} methods have also been used to explore models that incorporate both \gls*{ee} and \gls*{eph} interactions on an equal footing. Competition between distinct ordered phases favored by different interactions is common, with intermediate metallic phases or phase coexistence emerging as possible consequences~\cite{werner2007efficient, Fehske_2008, JohnstonDeveraux2013, Karakuzu2018, WangDemler2020, costa2020, Karakuzu2022stripe, Gotz2022, Ly2025, Costa2026antiferromagnetism}. It has also been argued that the intermediate metallic phase can have enhanced tendencies toward unconventional superconductivity~\cite{Nowadnick2015}. 

Here, we use \gls*{DQMC} to examine the interplay of \gls*{BOW} and \gls*{AFM} phases in the half-filled \gls*{SSHH} Hamiltonian under uniaxial strain. Prior work~\cite{Ly2025} on the unstrained model has shown that it  contains competing \gls*{BOW} and \gls*{AFM} phases, with a small region of potential coexistence between the two  (see also Fig.~\ref{fig:PD}). We show that introducing uniaxial strain, which leads to anisotropic hopping integrals $t_x > t_y$, tilts the competition in favor of bond order. We quantify the shift in the low-temperature ($\lambda, U)$ phase diagram previously obtained in the isotropic limit~\cite{Ly2025}.  We also evaluate the \gls*{BOW} transition temperature as a function of the strain and show that its evolution reflects a non-trivial competition between enhancement due to the fewer low-energy ordering possibilities, and suppression as the lattice becomes more quasi-one-dimensional. The \gls*{DQMC} methodology~\cite{blankenbecler81, white89a} is an especially potent approach for this problem since, at half-filling, particle-hole symmetry prevents the sign problem~\cite{loh90, troyer05, Mondaini2022} even when both the \gls*{ee} and \gls*{eph} interactions are present~\cite{Ly2025}.

This paper is organized as follows. Section~\ref{sec:methods} describes the (site) optical \gls*{SSHH} Hamiltonian, the \gls*{DQMC} method, and the various measurements we use to characterize the phases. Section~\ref{sec:results} presents our results.  We first describe the scaling of the \gls*{BOW} and \gls*{AFM} structure factors, which leads to the equilibrium phase diagram, along with local correlations whose evolution changes significantly at the phase boundaries.  We then present data for the spectral function, which exhibits clear pseudogap behavior characteristic of the \gls*{BOW} phase, including directionally dependent gap formation due to a combination of strain and \gls*{BOW} order. Finally, Sec.~\ref{sec:conclusions} summarizes our conclusions and points to open questions that remain.

\section{Model and Method}\label{sec:methods}

\subsection{The optical SSH-Hubbard Model} 
We study the optical \gls*{SSHH} model defined on a \gls*{2D} square lattice with periodic boundary conditions 
and lattice constant $a$. In this case, the lattice degrees of freedom are modeled using (quantum) harmonic oscillators on each site~\cite{Capone1997small}. The model's Hamiltonian is  
\begin{equation}\label{eq:ham}
    \begin{split}
    \mathcal{\hat H} =&- (t+\delta) \sum_{i, \sigma} 
 \left( 1 - \frac{\alpha}{t} (\hat{X}_{i}-\hat{X}_{i+\hat x}) \right)
  \left( \hat{c}^{\dagger}_{i, \sigma} \hat{c}^{\phantom\dagger}_{i+\hat x, \sigma}
 + {\mathrm{h.c.}} \right)\\
& -(t-\delta) \sum_{i, \sigma} \left( 1 -  \frac{\alpha}{t} (\hat{Y}_{i}-\hat{Y}_{i+\hat y}) \right)
  \left( \hat{c}^{\dagger}_{i, \sigma} \hat{c}^{\phantom\dagger}_{i+\hat y, \sigma}
 + {\mathrm{h.c.}} \right)\\
                 & + \sum_{i} \left(\frac{1}{2M} \hat{\boldsymbol{P}}_{i}^2 +
  \frac{M}{2} \, \Omega_{0}^2 \, \hat{\boldsymbol{R}}_{i}^2 \right)\\
                 & + U \sum_{i} \left(\hat{n}_{i\uparrow}-\frac{1}{2}\right)
  \left(\hat{n}_{i\downarrow}-\frac{1}{2}\right)
  - \mu \sum_{i, \sigma} \hat{n}_{i \sigma}. 
  \end{split}  
\end{equation}
Here, $\hat{c}_{i \sigma}$ ($\hat{c}_{i \sigma}^{\dagger}$) destroys (creates) an electron of spin $\sigma=\uparrow,\downarrow$ on site $i$, $\mu$ is the electron chemical potential, $t$ is the nearest-neighbor hopping in the absence of any strain, $\hat{n}_{i\sigma}=  \hat{c}^{\dagger}_{i \sigma} \hat{c}^{\phantom\dagger}_{i \sigma}$ is the number operator, and $U$ is the on-site Hubbard repulsion. 
The atomic displacement in the plane is described by the position operator $\hat{\boldsymbol{R}}_i = (\hat{X}_i,\hat{Y}_i)$ and associated momentum operators $\hat{\boldsymbol{P}}_i = (\hat{P}_{i,x},  \hat{P}_{i,y})$. The characteristic mass and frequency of the oscillators are $M$ and $\Omega_{0}$, respectively. The difference in displacements at the endpoints of a bond modulates the corresponding electronic hopping via an \gls*{eph} interaction $\alpha$.  {\it The key new ingredient of our study is the inclusion of strain.}  The hopping $t+\delta$ and $t-\delta$ are anisotropic, i.e.~different in the $\hat x$ and $\hat y$ directions. Note that even though $t_x \neq t_y$, the Fermi surface remains nested with momentum ${\bf Q}=(\pi/a,\pi/a)$, a key driving ingredient in \gls*{AFM} and \gls*{BOW} orders.

Throughout this work, we set our units such that $\hbar = t = M = k_\mathrm{B} = a=1$ and fix $\Omega_{0} = 0.5t$.  We will also report the magnitude of \gls*{eph} coupling using a dimensionless parameter $ \lambda=\frac{\alpha^2}{M \Omega_0^2 t} = \frac{8 \alpha^2}{M \Omega_0^2 W}$ with $W=8t$ the non-interacting band-width. Note, in this case, $\lambda$ is always defined with respect to the unstrained system and is independent of $\delta$. 

Our choice of strain modulates the `phonon-independent' hopping and the effective \gls*{eph} coupling $\alpha$. This is the physically appropriate form that naturally arises with strain when $\alpha$ is defined by the linear correction $ \alpha=\frac{dt}{dR}$ to the hopping parameter $t(R)$ due to the lattice displacements~\cite{bardeen1950deformation, harrison1956scattering}, where $R$ is the separation between neighboring (adjacent) sites. The formal procedure~\cite{slater1954simplified} to calculate $t$ is standard. Regardless of approximations used and the functional form of $t(R)$ (for example, an exponential~\cite{leman1962description} or power-law~\cite{harrison1989electronic}), the leading-order expression for $\alpha(R+\epsilon)$ is proportional to $t(R+\epsilon)$, where $\epsilon$ is the displacement caused by strain. 

\subsection{Determinant quantum Monte Carlo} 
We solved Eq.~\eqref{eq:ham} using the \gls*{DQMC} algorithm, as implemented in the \texttt{SmoQyDQMC.jl} package~\cite{SmoQyDQMC1, SmoQyDQMC2}. 

Formulated in the grand canonical ensemble at inverse temperature $\beta=\frac{1}{T}$, the \gls*{DQMC} algorithm evaluates thermal expectation values from the partition function $Z=\text{Tr} \big( e^{-\beta\mathcal{H}} \big)$. One introduces imaginary time slices $\Delta\tau=\beta/L$ with (discrete) \gls*{HS} $s=\{s_{i,l}\}$ and (continuous) phonon field configurations $x=\{x_{i,l}\}$, allowing for the analytic calculation of the trace
over the fermion operators
\begin{equation}
    \begin{split}
     Z\approx\sum_{s_{i,l}}\int &\prod_{i,l} dx_{i,l} \, e^{-S_\mathrm{B}(x)} \times \\
     &\prod_{\sigma=\uparrow,\downarrow}\det\left( I+\prod_{l=1}^L B_{\sigma}(s,l,x)\right).      
    \end{split}
\end{equation}
The $B_{\sigma}(s,l,x)$ are propagator matrices at a time slice $l$ for a particular instance of \gls*{HS} and phonon fields. $S_\mathrm{B}(x)$ is the purely bosonic contribution to the action; in the current model, it is given by the total action of $2L$  harmonic oscillators per site. The discretization (`Trotter') errors are $O(\Delta \tau^2)$ in the imaginary-time axis discretization constant $\Delta\tau$, and can be made less than the statistical sampling errors by using a sufficiently fine imaginary time grid.

In the simplest incarnation of \gls*{DQMC}, the \gls*{HS} and phonon fields are updated locally. That is, a change is proposed to a single $s_{i,l}$ or $x_{i,l}$ and is accepted or rejected according to the ratio of the fermion determinants before and after the move (as well as any changes in $S_\mathrm{B}$ when updating $x_{i,l}$).  However, much more efficient approaches have been developed for sampling phonon fields in general, and are needed in the case of the optical \gls*{SSH} interaction. In particular, \texttt{SmoQyDQMC.jl}~\cite{SmoQyDQMC1, SmoQyDQMC2} implements \gls*{HMC} updates to sample the phonon fields according to fictitious Hamiltonian dynamics. For a fixed \gls*{HS} field $s$, the total action is
\begin{equation}
    S(s,x)=S_\mathrm{B}(x)+S_\mathrm{F}(s,x)  \,\,.
\end{equation}
The fermionic contribution is 
\begin{equation}
    S_\mathrm{F}(x,s)=\sum_{\sigma=\uparrow,\downarrow}\log\left(\det{G_\sigma(\tau,\tau)}\right), 
\end{equation}
where $G_\sigma(\tau,\tau)$ is the equal-time Green's function for a given \gls*{HS} field configuration $s$ and phonon field configuration $x$. It is given in terms of the fermion determinant matrices 
\begin{equation}
    G_\sigma(\tau,\tau)=\Big(I+B_\sigma(\tau,0)B_\sigma(\beta,\tau) \Big)^{-1}
\end{equation}
with
\begin{equation}
    B_\sigma(\tau,\tau')=\prod_{l=(\tau/\Delta\tau)}^{(\tau'/\Delta\tau)} B_\sigma(s,l,x).
\end{equation}
To sample the phonons, a conjugate momentum $p$ is introduced for each phonon field $x$, such that an effective Hamiltonian
\begin{equation}
    H=S(s,x)+\frac{1}{2}p^2.
\end{equation}
is defined. 
We initially draw each $ p$ from a standard normal distribution. To obtain the proposed update to the phonon fields, the variables evolve via Hamiltonian dynamics for $N_t$ time steps to $p\rightarrow p^\prime,x\rightarrow x^\prime$ and are accepted with probability
\begin{equation}
    P=\min{(1,e^{-[H(x^\prime, p^\prime)-H(x, p)]})},
\end{equation}
making the update exact. In practice, a modified version of a standard \gls*{HMC} update called \gls*{EFA-HMC} is used to efficiently sample the phonon field; we refer the reader to Refs.~\cite{SmoQyDQMC1, Ostmeyer2025Minimal, Duane1987Hybrid, Beyl2018Revisiting} for more information. 

Importantly, particle-hole symmetry, which ensures that the up- and down-spin determinants have the same sign in the half-filled Hubbard model, is preserved for any choice of $t_{ij}$ as long as the lattice is bipartite.  As a consequence, the \gls*{SSHH} model has no sign problem at $\rho = \langle \hat n_{i\uparrow} + \hat n_{i\downarrow} \rangle = 1$.

In determining the equilibrium properties,
we ran \gls*{DQMC} simulations with fixed $\Delta\tau=0.05$. Additionally, the \gls*{EFA-HMC} updates were performed using $N_t=8$ time-steps with corresponding step-size $\Delta t = \pi/(2N_t)$. To obtain the ground-state phase diagram, we used $1\times10^4$ thermalization steps and $2\times10^4$ measurements and scaled the inverse temperature in proportion to the cluster size $\beta t=L$. For the finite-temperature transition, we used $2\times10^4$ thermalization steps and $4\times10^4$ measurements. After determining the ground-state phase diagram we ran additional \gls*{DQMC} simulations to determine the spectral and conductive properties across the phases for a strain of $\delta =0.3t$. For these calculations, we used lattice size of $12\times12$ with parameters $U=5t$, $\Omega_0=0.5t$ and $\beta=20/t$ over $\lambda\in[0.06,0.18]$ in increments of $0.02$. Each simulation used six independent walkers, $1.5\times 10^3$ thermalization steps, and $2\times10^3$ measurements.

\subsection{Observables} 
 
The \gls*{BOW} phases are most immediately revealed by the asymmetries in the hopping 
(kinetic energy) in the $x$ and $y$ directions
\begin{equation}
    K_{x(y)} \equiv \frac{1}{N}\sum_i\langle \hat K_{i}^{x \left( y \right)} \rangle = \frac{1}{2N}\sum_{i,\sigma}\langle \hat{c}_{i,\sigma}^{\dagger} \hat{c}^{\phantom{\dagger}}_{i + \hat{x} \left( \hat{y} \right),\sigma} + \mathrm{h.c.} \rangle 
\end{equation} 
and the phonon displacements in the $\hat x$ and $\hat y$ directions. We obtain the former by averaging appropriate matrix elements of the equal-time Green's function. In the isotropic $t_x=t_y$ limit,  $\langle \hat{K}^x_{i} \rangle  = \langle \hat{K}^y_{i} \rangle $ in the unordered phase while the external strain $t_x \neq t_y$ breaks $C_4$ symmetry.  Nevertheless, the average kinetic energy shows kinks in its behavior when \gls*{BOW} order sets in. Meanwhile, the phonon displacements, which are simple averages of the phonon coordinates being sampled, always obey $\langle \hat X_{i} \rangle = \langle \hat Y_{i} \rangle =0 $ but can have distinct \gls*{rms} values  ${\cal X}_\mathrm{rms}=\sqrt{\langle\hat{X}_i^2\rangle}$  and $ {\cal Y}_\mathrm{rms}=\sqrt{\langle\hat{Y}_i^2\rangle}$ in the \gls*{BOW} phase. The fluctuations in the lattice displacement thus offer clear signals of phase transitions through kinks in their evolution with temperature $T$ or \gls*{eph} coupling $\lambda$. In addition, we examine the double occupancy, $D = \langle \hat{n}_{i\uparrow}  \hat{n}_{i\downarrow} \rangle$, a fundamental probe of the effects of the on-site repulsion $U$, including local moment formation.

In addition to these local quantities, we also measured the equal-time spin-spin  
\begin{equation}
c_{\rm spin}(\boldsymbol{r}) = \sum_i \langle \hat S_{i}^{x} \hat S_{i + \boldsymbol{r}}^{x} \rangle
\end{equation}
and bond-bond 
\begin{equation}
c^{x\left(y\right)}_{\rm bow}(\boldsymbol{r}) = \sum_i \langle \hat{K}^{x \left( y \right)}_i \hat{K}^{x \left( y \right)}_{i + \boldsymbol{ r}}\rangle 
\end{equation}
correlation functions in real-space, where $S_{i}^{x} = \frac{1}{2}  \left(\hat c_{i\uparrow}^{\dagger} \hat c_{i\downarrow}^{\phantom{\dagger}} + \hat c_{i\downarrow}^{\dagger} \hat c_{i\uparrow}^{\phantom{\dagger}}\right)$. These quantities probe the \gls*{AFM} and \gls*{BOW} correlations at separation $\boldsymbol{r}$. Expectation values of such four-fermion operator observables are obtained via Wick's theorem from the two-fermion Green's functions. 

The \gls*{AFM} and \gls*{BOW} structure factors at $\boldsymbol{Q}=(\pi,\pi)$ are obtained by Fourier transform  $S_{\mathrm{AFM}}= \frac{1}{N}\sum_{\boldsymbol{r}} e^{\mathrm{i} \boldsymbol{Q}\cdot\boldsymbol{r}} 
c_{\rm spin}(\boldsymbol{r})$, and $S_{{\rm BOW}}^{x \left( y \right)} = \frac{1}{N}\sum_{\boldsymbol{r}} e^{\mathrm{i} \boldsymbol{Q}\cdot\boldsymbol{r}}
c^{x\left(y\right)}_{\rm bow}(\boldsymbol{r})$. Throughout this work, we distinguish the $x$ and $y$ directions due to the hopping anisotropy. In the simplest picture, the structure factors are expected to be independent of lattice size in disordered phases where the real-space correlation functions fall exponentially with distance and hence do not contribute to the sum over displacements when $r>\xi$, the correlation length.  In contrast, the structure factors grow linearly with the number of lattice sites $N$ when long-range order is present, and contributions to the sum occur for all $r$ up to the linear lattice size $L$. A more careful analysis that isolates the $O(L)$ corrections will be done with finite-size scaling in what follows.

The invariant correlation ratios are defined by 
\begin{equation}\label{eq:Rgamma}
 R_\gamma(L) = 1 - \frac{S_{\gamma}(\boldsymbol{Q}+\Delta\boldsymbol{Q}_L)}{S_{\gamma}(\boldsymbol{Q})}
\end{equation}
$(\gamma={\rm BOW,AFM})$
with $\boldsymbol{Q}=(\pi,\pi)$ the ordering wave vector and $\Delta\boldsymbol{Q} = (\frac{2 \pi}{L},0)$, $(0,\frac{2 \pi}{L})$. These quantities can be used to accurately determine the location of critical points in the thermodynamic limit. When evaluated on lattices of different linear size $L$, and at low temperatures, $R_\gamma(L)$ exhibits crossings at the phase boundary.

We can extract the fermionic quantities from the equal-time Green's functions. If we measure the imaginary-time-dependent Green's function $G_\sigma(\boldsymbol{k},\tau)$, we can also probe the dynamics.  Specifically, the spectral function $A_\sigma(\boldsymbol{k},\omega)$ obeys the integral equation 
 \begin{align}
 G_\sigma(\boldsymbol{k},\tau)=  \int_{-\infty}^{\infty} d\omega \frac{e^{-\tau \omega}}{1+e^{-\beta \omega}} A_\sigma(\boldsymbol{k},\omega) \label{eq:Aw}
 \end{align}
where
\begin{align}
G_{\sigma}\left(\boldsymbol{k},\tau\right)=\frac{1}{N}\sum_{i,j}e^{-{\rm i}\boldsymbol{k}\cdot\left(\boldsymbol{r}_{i}-\boldsymbol{r}_{j}\right)}\left\langle \hat{c}_{i,\sigma}^{\phantom{\dagger}}(\tau)\hat{c}_{j,\sigma}^{\dagger}(0)\right\rangle.
\end{align} The density of states can also be accessed using the position-space Green's function as $G_{\sigma}(\boldsymbol{r}=0,\tau)=\frac{1}{N}\sum_{\boldsymbol{k}}G_\sigma(\boldsymbol{k},\tau)$. Since our Hamiltonian observables are spin-independent, we will drop the $\sigma$ subscript henceforth. 

Similarly, we obtained the optical conductivity from the current-current correlation function
\begin{equation}\label{eq:currentcurrent}
    \Lambda_{\alpha,\beta}\left(\boldsymbol{r},\tau\right)=\frac{1}{N}\sum_{i}\left\langle \hat{J}^\alpha(\boldsymbol{r}+\boldsymbol{r}_i,\tau)\hat{J}^\beta(\boldsymbol{r}_i,0)\right\rangle,
\end{equation}
where the current operator for direction $\alpha$ is given by
\begin{equation}\label{eq:current}
    \hat{J}^\alpha(\boldsymbol{r}_i)=-\mathrm{i}t_{i,\alpha}^{\rm eff}\sum_\sigma \left(c_{i+\alpha,\sigma}^{\dagger}c_{i,\sigma}^{\phantom{\dagger}}-c_{i,\sigma}^{\dagger}c_{i+\alpha,\sigma}^{\phantom{\dagger}}\right)
\end{equation}
and $\alpha$ and $\beta$ $\in\{\hat{x},\hat{y}\}$, 
$t_{i,\hat{x}}^{\rm eff}=(t+\delta)[1-\tfrac{\alpha}{t}\left(X_i-X_{i+\hat{x}}\right)]$, and $t_{i,\hat{y}}^{\rm eff}=(t-\delta)[1-\tfrac{\alpha}{t}\left(Y_i-Y_{i+\hat{y}}\right)]$. $\Lambda\left(\boldsymbol{r}_i,\tau\right)$ is related to the optical conductivity tensor $\sigma(\omega)$ via the relation
\begin{equation}\label{eq:ACboson}
    \Lambda(\boldsymbol{q}=0, \tau) = \int_{0}^{\infty}d\omega\frac{\omega\left( e^{-\tau\omega}+e^{-(\beta-\tau)\omega}\right)}{1-e^{-\beta\omega}}\sigma(\omega). 
\end{equation}

Inverting Eqs.~\eqref{eq:Aw} and~\eqref{eq:ACboson} is 
an ill-posed \gls*{AC} problem, and multiple methods exist for performing this task on noisy data such as that obtained from \gls*{DQMC}. We inverted 
Eq.~\eqref{eq:Aw} using the \texttt{SmoQyDEAC.jl} package~\cite{Neuhaus2024, Neuhaus2024-r1.1}, which implements the \gls*{DEAC} algorithm~\cite{Nichols2022Parameter}. \gls*{DEAC} utilizes a population of candidate spectral functions which stochastically mix to converge on a likely candidate spectrum. Unlike the commonly used \gls*{MaxEnt} method~\cite{Jarrell1996Bayesian}, \gls*{DEAC} does not require a user-supplied model function. It has also been shown to improve resolution of sharp features in fermionic correlation functions~\cite{Neuhaus2024}. \texttt{SmoQyDEAC.jl} was run over 10,000 instances to generate spectra. \gls*{DEAC} is less reliable for analytically continuing bosonic correlation functions. We, therefore, used the \gls*{MaxEnt} method as implemented in the ana\_cont package~\cite{Jarrell1996Bayesian, Kaufmann2023anacont} to invert Eq.~\eqref{eq:ACboson}. In this case, we used a Gaussian centered at $\omega=0$ with a width of $8t$ and normalized to an integral of $1.0$ as the default function.

 \begin{figure}[t!]
    \centering
    \includegraphics[width=\columnwidth]{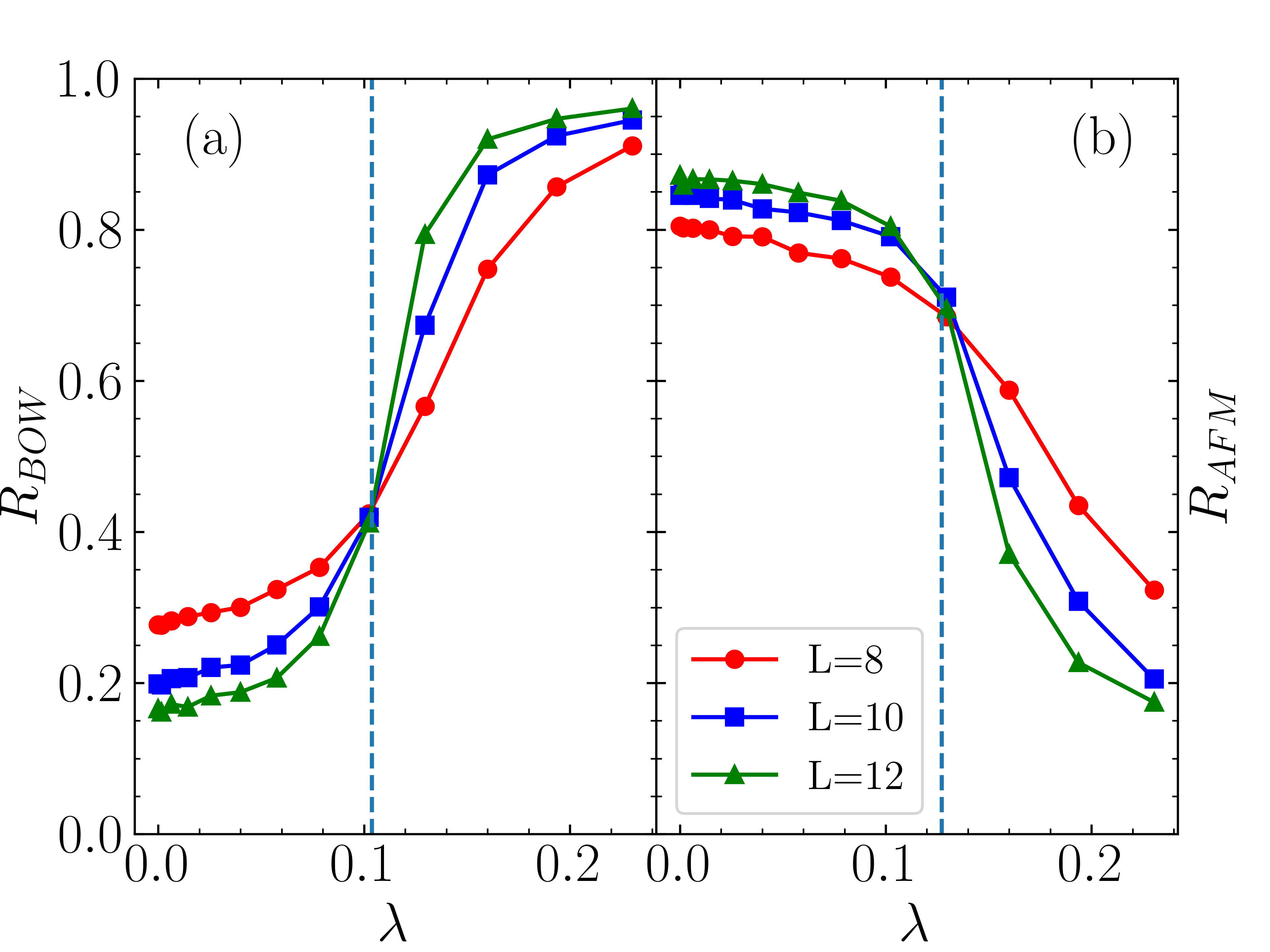}
    \caption{Correlation ratios for the (a) \gls*{BOW} and (b) \gls*{AFM} phases for $\delta=0.3t$, $U=5t$, and $\Omega_0=0.5t$.  The critical value of the dimensionless coupling $\lambda_\mathrm{c}$ in the thermodynamic limit is obtained from the intersection of the interpolated curves for the different lattice sizes $L=\beta t=8,10,12$, as indicated by the dashed lines.}
    \label{fig:Rvslambda}
\end{figure}

 \begin{figure}[htpb]
    \centering
    \includegraphics[width=\columnwidth]{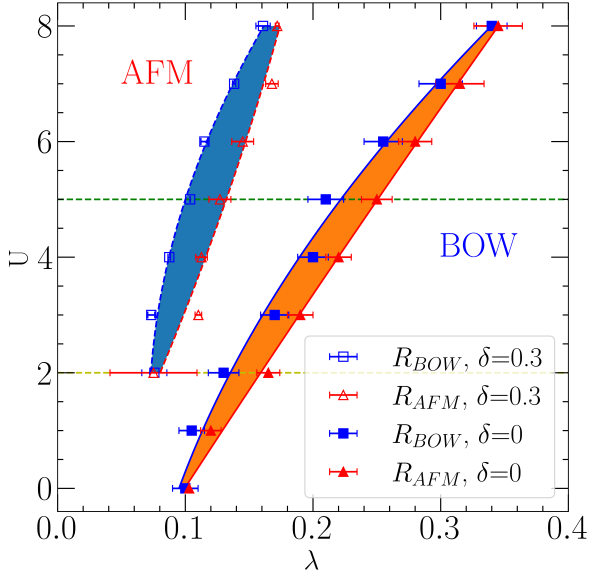}
    \caption{Ground state phase diagram of the optical \gls*{SSHH} model at half-filling in the plane of \gls*{eph} coupling and on-site repulsion $U$. Here phonon  frequency $\Omega_0=0.5t$ and strain  $\delta=0.3t$. Blue open triangles give the \gls*{BOW} phase boundaries and red open triangles give the \gls*{AFM} phase boundaries. We show results for the unstrained case, $\delta=0$, (from \cite{Ly2025}) for comparison (filled symbols). The shaded regions indicate coexistence. Figures \ref{fig:Rvslambda} and \ref{fig:localvslambda} were obtained using data from the line cuts at $U=5t$ (green dashed line) and $U=2t$ (yellow dashed line), respectively.}
    \label{fig:PD}
\end{figure}

\section{Results}\label{sec:results}

\subsection{Low temperature phase diagram}
We first address the low-temperature orders. Figure \ref{fig:Rvslambda} plots the invariant correlation ratios $R_{\rm AFM}$ and $R_{\rm BOW}$ as functions of the dimensionless coupling strength $\lambda$ for fixed on-site repulsion $U=5t$ and phonon frequency $\Omega_0=0.5t$. The hopping anisotropy is $\delta = 0.3t$. $R_{\rm BOW}(L)$ increases with $\lambda$, and curves for different $L$ cross 
at $\lambda \sim 0.104$, signaling the entry into a \gls*{BOW} phase. On the other hand, $R_{\rm AFM}(L)$ decreases with $\lambda$, and its curves for different $L$ cross at $\lambda \sim 0.127$ as \gls*{AFM} order is lost.  Importantly, the two critical $\lambda$ values do not coincide, suggesting a narrow region where both orders coexist, similar to reports for the unstrained model~\cite{Ly2025}. 

Performing such $\lambda$ sweeps for different fixed values of $U$,  we can construct the ground state phase diagram of the site \gls*{SSHH} model in the ($\lambda, U$) plane shown in Fig.~\ref{fig:PD}. Here, \gls*{AFM} order dominates when the on-site $U$ is large, while \gls*{BOW} order forms as $\lambda$ increases. The shaded region shows the coexistence region. We present results for the unstrained isotropic ($\delta=0$) case for comparison~\cite{Ly2025}. For both the strained and unstrained cases, the nesting at $\boldsymbol{Q}=(\pi,\pi)$ causes ground state \gls*{AFM} order for any $U$ when $\lambda \rightarrow 0$ and a finite $\lambda$ is required to stabilize \gls*{BOW} order. Crucially, applying strain drives the \gls*{AFM} and \gls*{BOW} phase boundaries to smaller $\lambda$, reflecting an increased tendency towards ground state \gls*{BOW} order. In the isotropic case, a \gls*{BOW} can form in any of four degenerate patterns: in either of the $x$ or $y$ directions and then, once a spatial direction is selected, in either of two sets of alternating bonds. When $\delta > 0$ one of these degeneracies is lifted and the \gls*{BOW} prefers to form in the $x$ ($y$) direction when $t_x = 1 + \delta > t_y = 1 - \delta$ ($t_y = 1 + \delta > t_x = 1 - \delta$). The phase boundary also has a steeper slope in the strained case $\delta=0.3t$; when $\delta \neq 0$  the critical value of $\lambda$ needed to induce \gls*{BOW} order is considerably less sensitive to $U$.  This behavior plausibly follows from including a $U$-independent hopping anisotropy. Interestingly, the width of the coexistence region does not appear to decrease when the system is strained.

\begin{figure}[t!]
    \centering
    \includegraphics[width=\columnwidth]{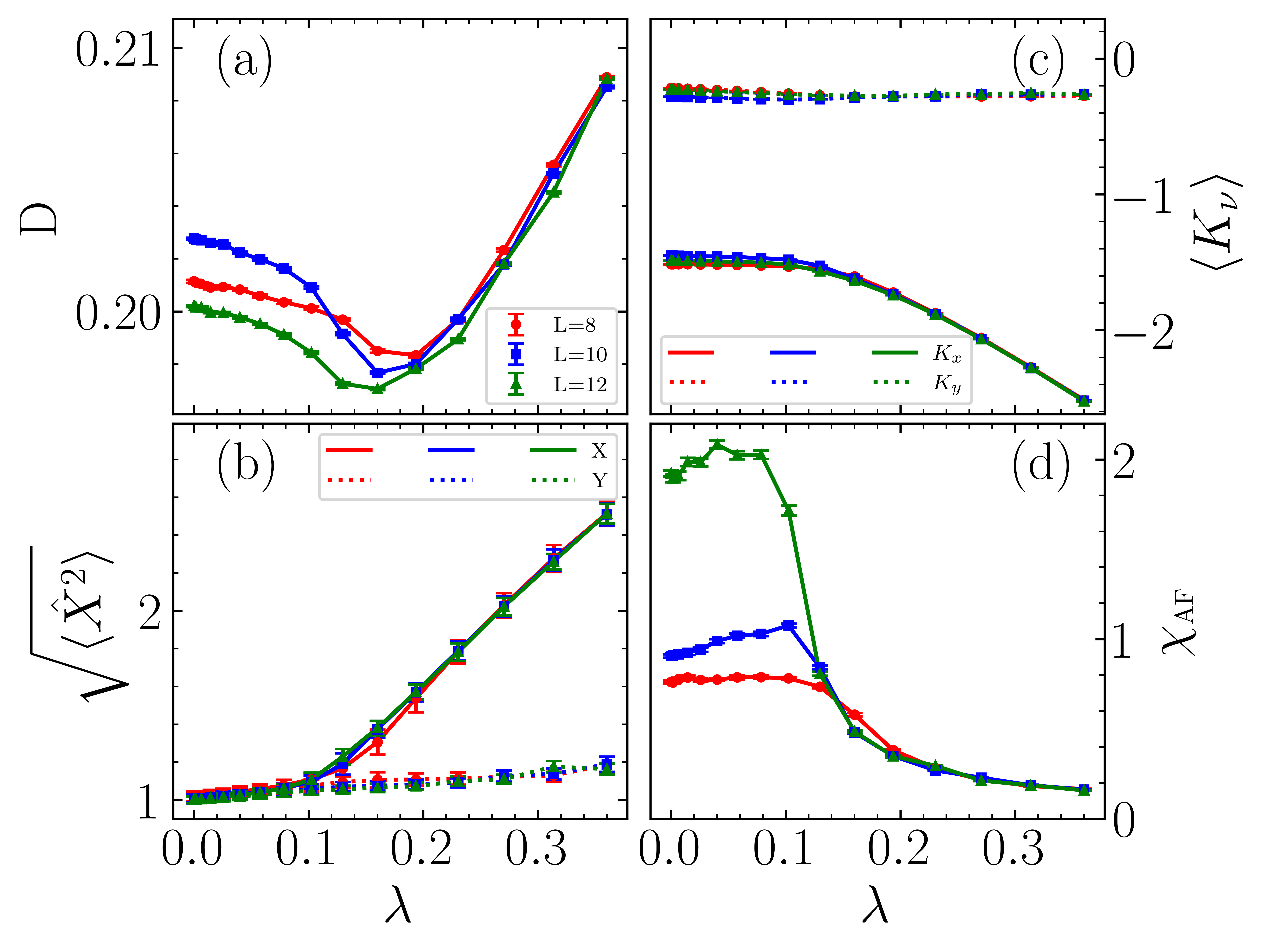}
    \caption{Local observables as a function of the \gls*{eph} coupling strength for fixed $\delta=0.3t$, $U=2t$, and $\Omega_0=0.5t$. Results are shown for the average values of (a) the double occupancy $D$; (b) the root mean square displacements ${\cal X}_\mathrm{rms}$ and ${\cal Y}_\mathrm{rms}$; (c) the direction-resolved kinetic energies $K_x$ and $K_y$, and the \gls*{AFM} susceptibility $\chi_\text{AFM}$. Results are also shown for different lattice sizes $L$ as indicated by the legend in panel (a). Here, we find that the $L$ dependence is weak, as is often the case for local observables. Panel (d) shows an additional global observable-- the \gls*{AFM} susceptibility, $\chi_{\rm AFM}$.}
    \label{fig:localvslambda}
\end{figure}

Figure~\ref{fig:localvslambda} plots the evolution of several local observables as a function of $\lambda$. These quantities further support the phase diagram in Fig.~\ref{fig:PD} and provide additional insight into the ordered phases. The double occupancy $D$ in Fig.~\ref{fig:localvslambda}(a) shows a shallow minimum near the AFM-BOW transition. In contrast, both the directionally resolved kinetic energy and \gls*{rms} lattice displacements show clear break-away behavior. Fig.~\ref{fig:localvslambda}(b) plots the $\mathcal{X}_\mathrm{rms}$ and $\mathcal{Y}_\mathrm{rms}$ displacements; they are degenerate in the \gls*{AFM} phase but then split at the onset of \gls*{BOW} order, with $\mathcal{X}_\mathrm{rms}$ increasing much faster than $\mathcal{Y}_\mathrm{rms}$. Similarly, Fig.~\ref{fig:localvslambda}(c) plots the average kinetic energies $K_x$ and $K_y$. These quantities are already unequal in the \gls*{AFM} phase at small $\lambda$ 
because $\delta \ne 0$; however, $K_x$ shows a clear change in behavior upon entry into the \gls*{BOW},
transitioning from being essentially $\lambda$ independent to growing rapidly in magnitude. 

Finally,  Fig.~\ref{fig:localvslambda}(d) shows an additional global observable, the \gls*{AFM} susceptibility, $\chi_{\rm AFM}$. While the structure factors defined in Sec.~IIC involve sums of equal-time spin or kinetic-energy correlation functions at different {\it spatial} separations, the associated susceptibilities include an additional integral over {\it imaginary} time separation $\tau$. $\chi_{\rm AFM}$ shows a sharp drop upon loss of magnetic order as $\lambda$ increases. It changes from a behavior characteristic of long-range spatial order (when the correlation length $\xi_{\rm AFM}$ exceeds the linear lattice size $L$, $\chi_{\rm AFM} \propto N$) to being lattice-size independent, indicating a finite correlation length $\xi_{\rm AFM} < L$.

 \begin{figure}[t]
    \centering
    \includegraphics[width=\columnwidth]{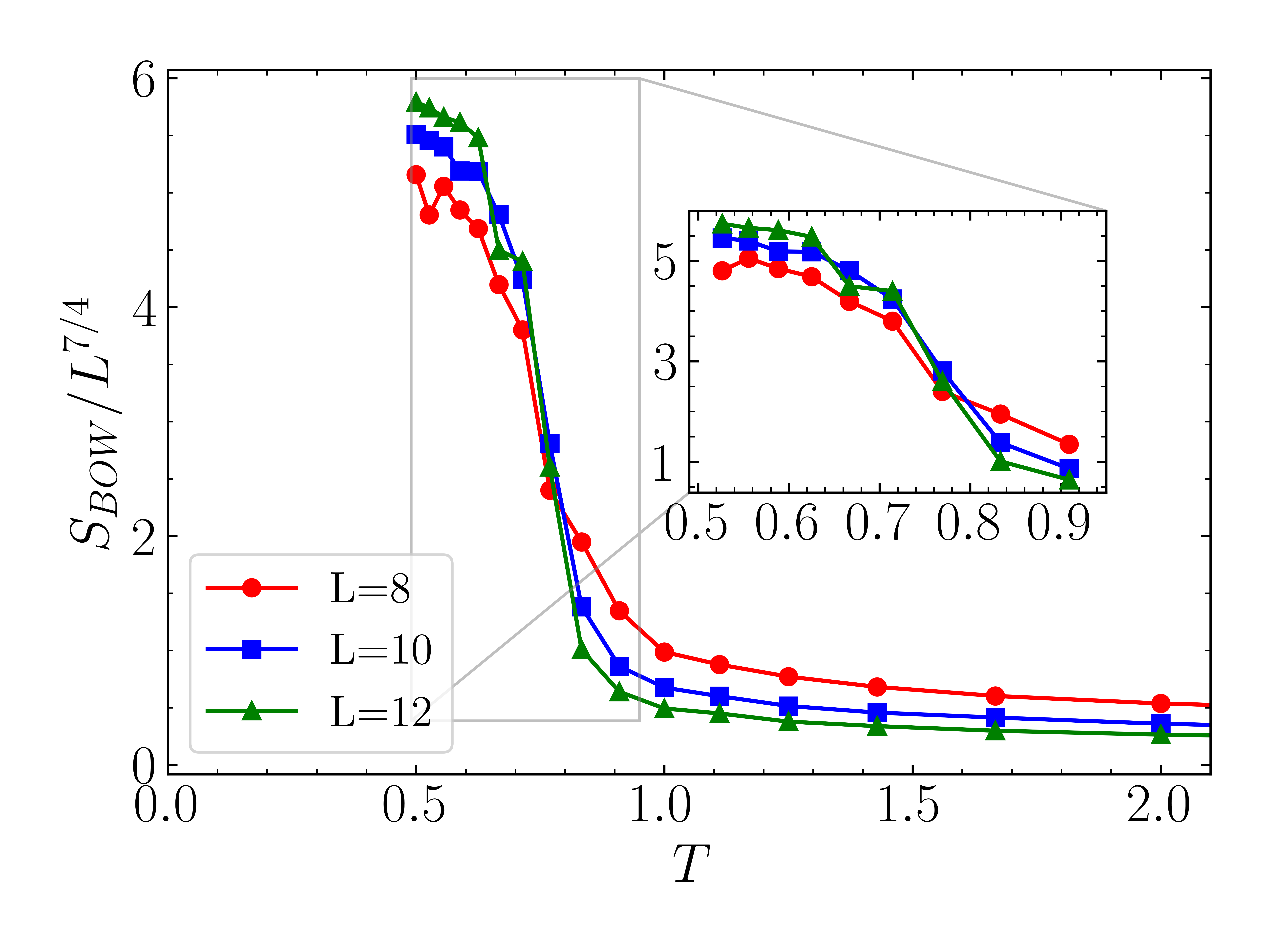}
    \caption{Crossing of the rescaled structure factors for the \gls*{BOW} phases for $U=4t$, $\Omega_0=t$, and $\lambda=1$ for the strained $\delta=0.3t$ case. We use the exponent ratio $\gamma/\nu = 7/4$, which is shared by the Ashkin-Teller, four-state-Potts, and Ising universality classes in \gls*{2D}.}
    \label{fig:SscalevsT}
\end{figure}

 \begin{figure}[t]
    \centering
    \includegraphics[height=3.0in,width=3.0in]{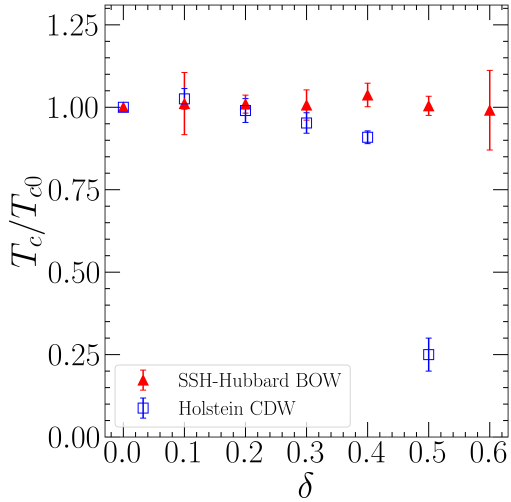}
    \caption{A plot of $T_\mathrm{c}(\delta)$, normalized to the isotropic value  $T_{c0} \equiv T_\mathrm{c}(\delta=0)$, obtained from the \gls*{BOW} correlation ratio crossings of $L=8$, $10$ and $12$, for $U=4t$, $\Omega_0=t$, and $\lambda=1$. For comparison, results for the strained Holstein case ($U=0$) are also given~\cite{Cohen-Stead2019}. The values of $T_{c0}$ are $T_{c0} \sim 0.17$ (CDW transition) for Holstein and $T_{c0} \sim 0.78$ for \gls*{SSHH} (BOW transition).}
    \label{fig:betacvsdelta}
\end{figure}

 \begin{figure}[t]
    \centering
    \includegraphics[width=\columnwidth]{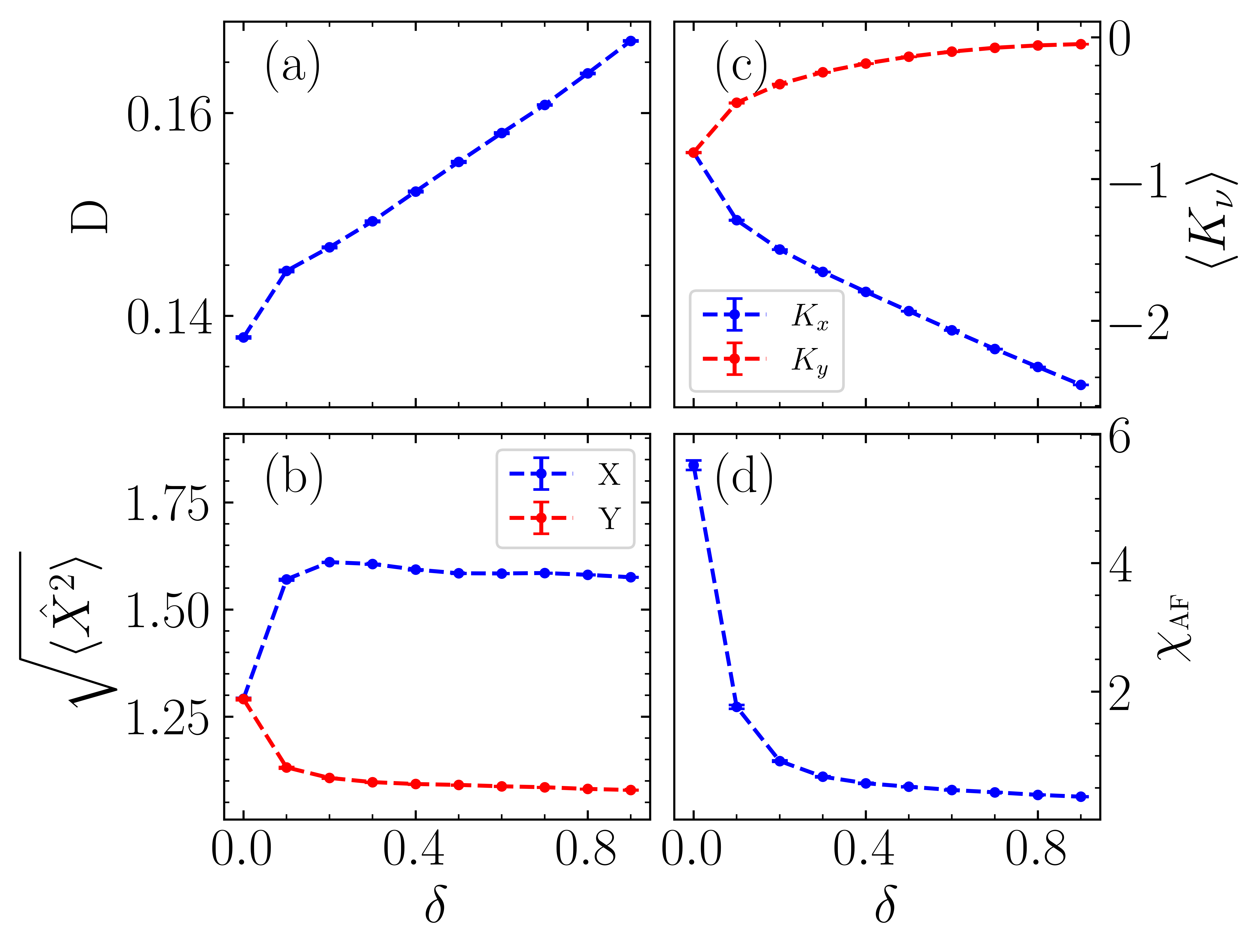}
    \caption{$D, K_\nu, \sqrt{(X_\nu^2)},$ and $\chi_{\rm AFM}$ for $U=4t$, $\Omega_0=0.5t$, $\lambda=0.21$, $L=\beta t=8$ and changing $\delta$ (horizontal axis). For these parameters, and for $\delta=0.3t$, the system is in the \gls*{BOW} phase. See the phase diagram of Fig.~\ref{fig:PD}.}
    \label{fig:localvsdeltalam0p21}
\end{figure}

 \begin{figure}[t]
    \centering
    \includegraphics[width=\columnwidth]{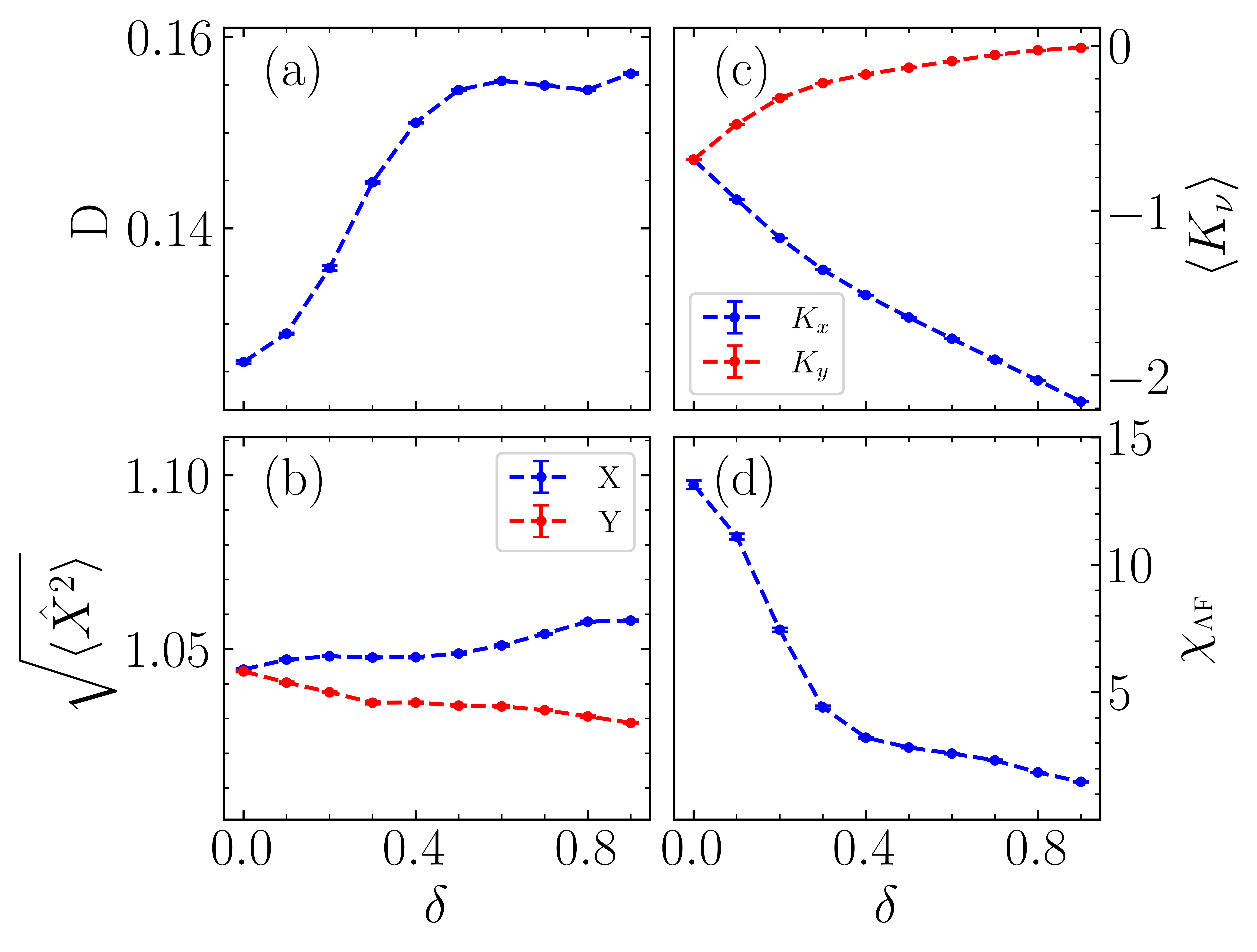}
    \caption{$D, K_\nu, \sqrt{(X_\nu^2)},$ and $\chi_{\rm AFM}$ for $U=4t$, $\Omega_0=0.5t$, $\lambda=0.04$, $L=\beta t=8$ and changing $\delta$ (horizontal axis). For these parameters, and for $\delta=0.3t$, the system is in the \gls*{AFM} phase. See the phase diagram of Fig.~\ref{fig:PD}.}
    \label{fig:localvsdeltalam0p04}
\end{figure}
\subsection{The finite $T$ \gls*{BOW} transition}

We turn now to the finite temperature transition into the \gls*{BOW} phase. As discussed earlier, the structure factor is independent of lattice size in the disordered phase, while it is proportional to $L^2$ in the ordered phase. Precisely at the critical point there is a non-trivial behavior: the structure factor, when scaled by $L^{-\gamma/\nu}$, is independent of $L$. Thus plots of $S_{\rm BOW}/L^{\gamma/\nu} $ versus $T$ will have a universal crossing at $T_\mathrm{c}$. Figure \ref{fig:SscalevsT} illustrates this behavior, and yields the value $T_\mathrm{c} \sim 0.78t$ ($\beta_c\sim 1.28/t$) after adopting $\gamma/\nu = 7/4$. In the absence of strain, the \gls*{BOW} phase has four symmetry-related ground states associated with the $C_4$ lattice symmetry and translations of the square lattice, and is represented by a ${\mathcal Z}_4$-type \gls*{BOW} order parameter. This discrete lattice symmetry is distinct from the $O(4)$ symmetry of the optical-\gls*{SSH} Hamiltonian ($U = 0$)~\cite{silva2025twodimensional}, which reflects an internal electronic symmetry. On its own, the fourfold \gls*{BOW} degeneracy is insufficient to fix finite-temperature universality class. In particular, ${\cal Z}_4$-symmetric systems can exhibit Ashkin-Teller criticality~\cite{deFelicio2018}, whose critical exponents vary continuously between decoupled Ising and four-state-Potts limits with fixed $\gamma/\nu = 7/4$. The four-state Potts point is a limiting case with $\gamma=7/6$ and $\nu=2/3$. Conversely, with the application of strain, we expect the \gls*{BOW} transition to instead fall under the \gls*{2D} Ising (${\cal Z}_2$) universality class with $\gamma=7/4$ and $\nu=1$. Importantly, the analysis in Fig.~\ref{fig:SscalevsT} cannot distinguish between these cases as all depend on the shared ratio $\gamma/\nu = 7/4$.  

\subsection{Dependence on the strain-induced anisotropy $\delta$ }

Figure~\ref{fig:betacvsdelta} shows the $\delta$ dependence of $T_\mathrm{c}$ 
for $U = 4t$, $\Omega_0 = t$, and $\lambda = 1$. The value of $T_\mathrm{c}(\delta=0.3t)$ is within the error bars of the unstrained critical point~\cite{Ly2025}. The trajectory is flat for $0 < \delta < 0.6t$: the critical temperature is unaltered by strain within error bars. In comparison, in the Holstein case, $T_c$ shows a noticeable decrease from $T_\mathrm{c}  \sim  0.17t$ to $T_\mathrm{c} \sim  0.15t$, when $\delta$ grows from $\delta=0$ to $\delta=0.4t$.  The critical temperature then falls even more to $T_\mathrm{c} \sim 0.04t$ at $\delta=0.5t$~\cite{Cohen-Stead2019}. The relative insensitivity of $T_\mathrm{c}$ to strain for the \gls*{BOW} phase in the \gls*{SSHH} model that we studied, as compared to the \gls*{CDW} phase in the Holstein model, can be attributed to the dependence of the \gls*{eph} coupling energy scale on the non-interacting bandwidth. In the Holstein model, the coupling strength is independent of strain while in the \gls*{SSHH} model the coupling strength scales directly with $t+\delta$.\footnote{Note that our \gls*{SSH} results are for $U=4t$ but the Holstein values are at $U=0$.}

Figures~\ref{fig:localvsdeltalam0p21} (in the \gls*{BOW} phase) and
\ref{fig:localvsdeltalam0p04} (in the \gls*{AFM} phase) provide further data for the $\delta$ dependence at low temperature, showing changes in the same observables
whose evolutions with $\lambda$ were tracked in  
Fig.~\ref{fig:localvslambda}. As expected, in both cases the disparities between the kinetic energies and the \gls*{rms} phonon displacements grow with increasing hopping anisotropy; however, in the \gls*{AFM} phase (Fig.~\ref{fig:localvsdeltalam0p04}) the \gls*{rms} phonon displacement differences are only a few percent over the entire range of $\delta$ but are almost a factor of two in the \gls*{BOW} phase (Fig.~\ref{fig:localvsdeltalam0p21}). In both cases, ${K}_y \rightarrow 0$ as $\delta \rightarrow 1$, when the lattice breaks into independent chains in the $x$ direction. Finally, strain systematically weakens the \gls*{AFM} order [see Fig.~\ref{fig:localvsdeltalam0p21}(d)].

\subsection{Dynamical correlation functions}

\begin{figure}[t]
    \centering
    \includegraphics[width=\linewidth]{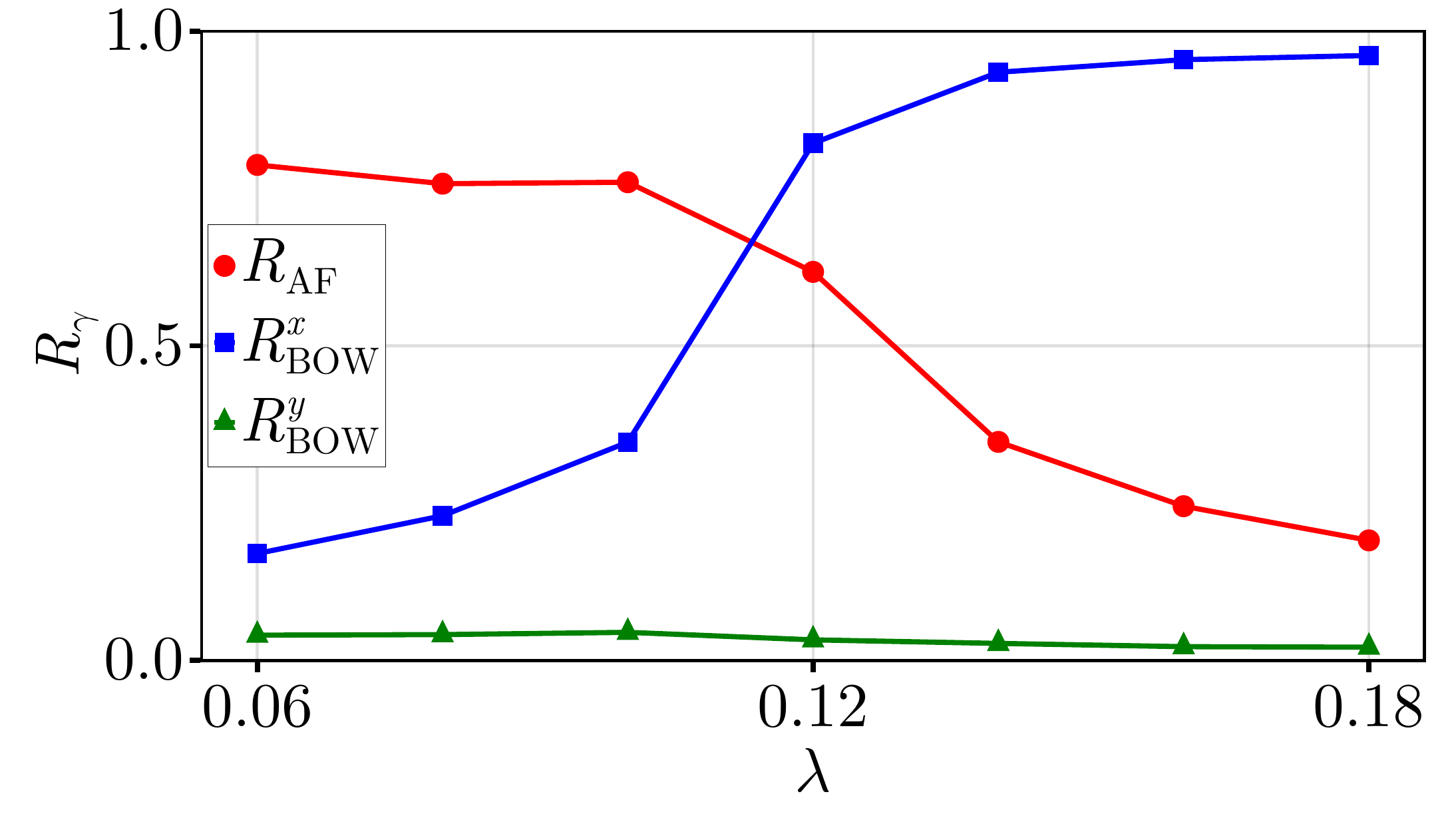}
    \caption{$R_{\rm AFM}(\pi,\pi)$ and $R_{\rm BOW}^{\rm x,y}(\pi,\pi)$ correlation ratios for $\delta =0.3$, $U=5$, $\Omega_0=0.5$ and $\beta=20$ over $\lambda\in[0.06,0.18]$ on a $12\times12$ lattice. 
    }
    \label{fig:Rgamma}
\end{figure}

Figure~\ref{fig:Rgamma} plots the correlation ratios of Eq.~\eqref{eq:Rgamma} for the finite temperature of $\beta =20/t$ with the parameters $\delta =0.3t$, $U=5t$, and $\Omega_0=0.5t$ on a $12\times12$ lattice over a range of \gls*{eph} coupling strengths. As expected, we do not see any signal of a \gls*{BOW} phase in the unstrained ($y$) direction of the lattice; however, there is a clear ordering into a $(\pi,\pi)$ \gls*{BOW} phase along the strained $(x)$ direction above $\lambda=0.1$. This coupling value is similar to that obtained at zero temperature using a scaling analysis, suggesting that $\beta = 20/t$ is low enough to reflect the model's ground-state properties for these parameters. 

\begin{figure}[t]
    \centering
    \includegraphics[width=\linewidth]{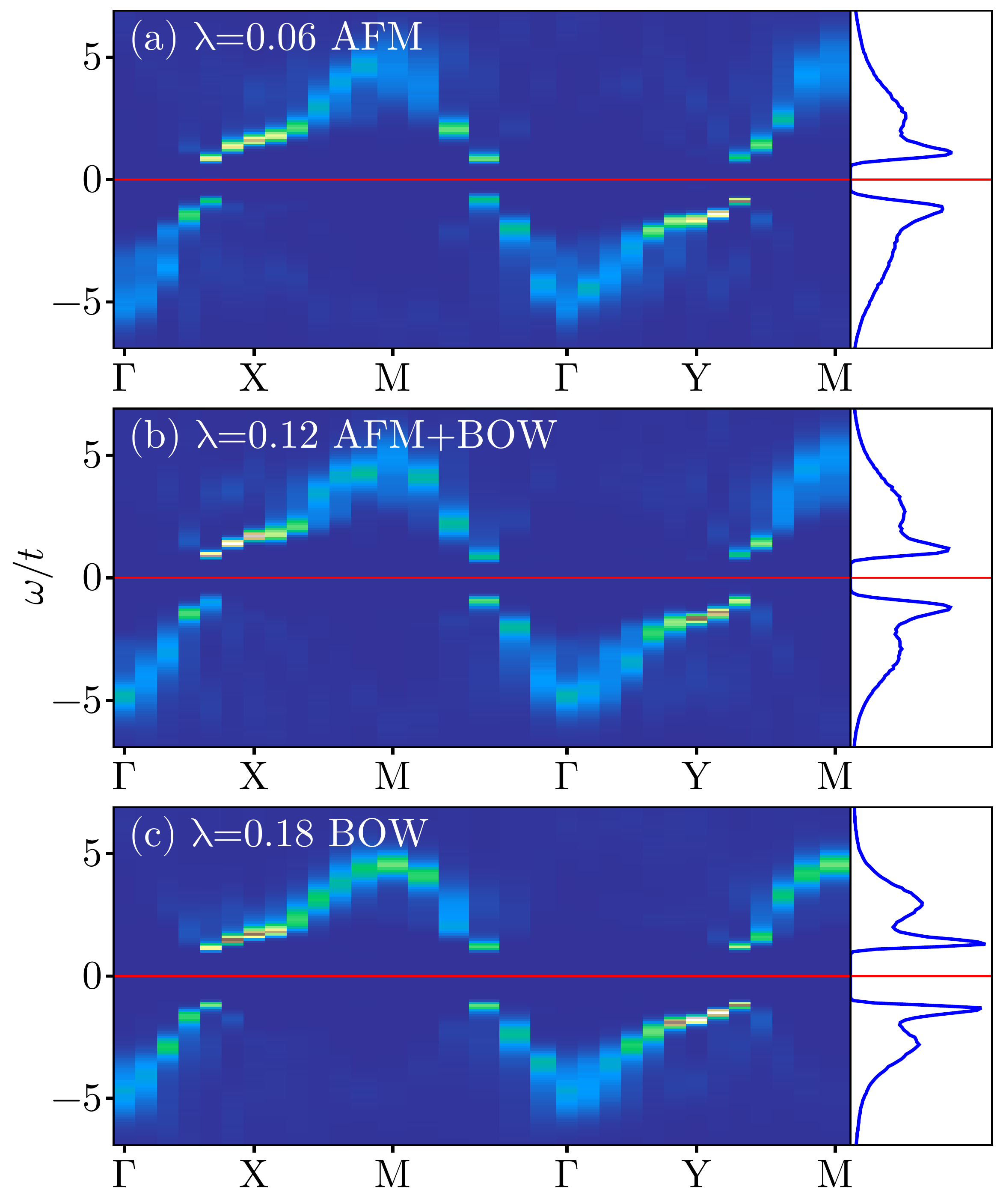}
    \caption{Low-temperature ($\beta=20/t$) Spectral functions plotted along the high-symmetry cuts of the first Brillouin zone. These results were obtained on 
    $12\times12$ clusters with fixed $\delta =0.3t$, $U=5t$, and $\Omega_0=0.5t$ and (a) in the \gls*{AFM} phase ($\lambda = 0.06$), (b) coexistence region ($\lambda = 0.12$), and (c) \gls*{BOW} phase ($\lambda = 0.18$).}
    \label{fig:Aw}
\end{figure}

\begin{figure}[t]
    \centering
    \includegraphics[width=\linewidth]{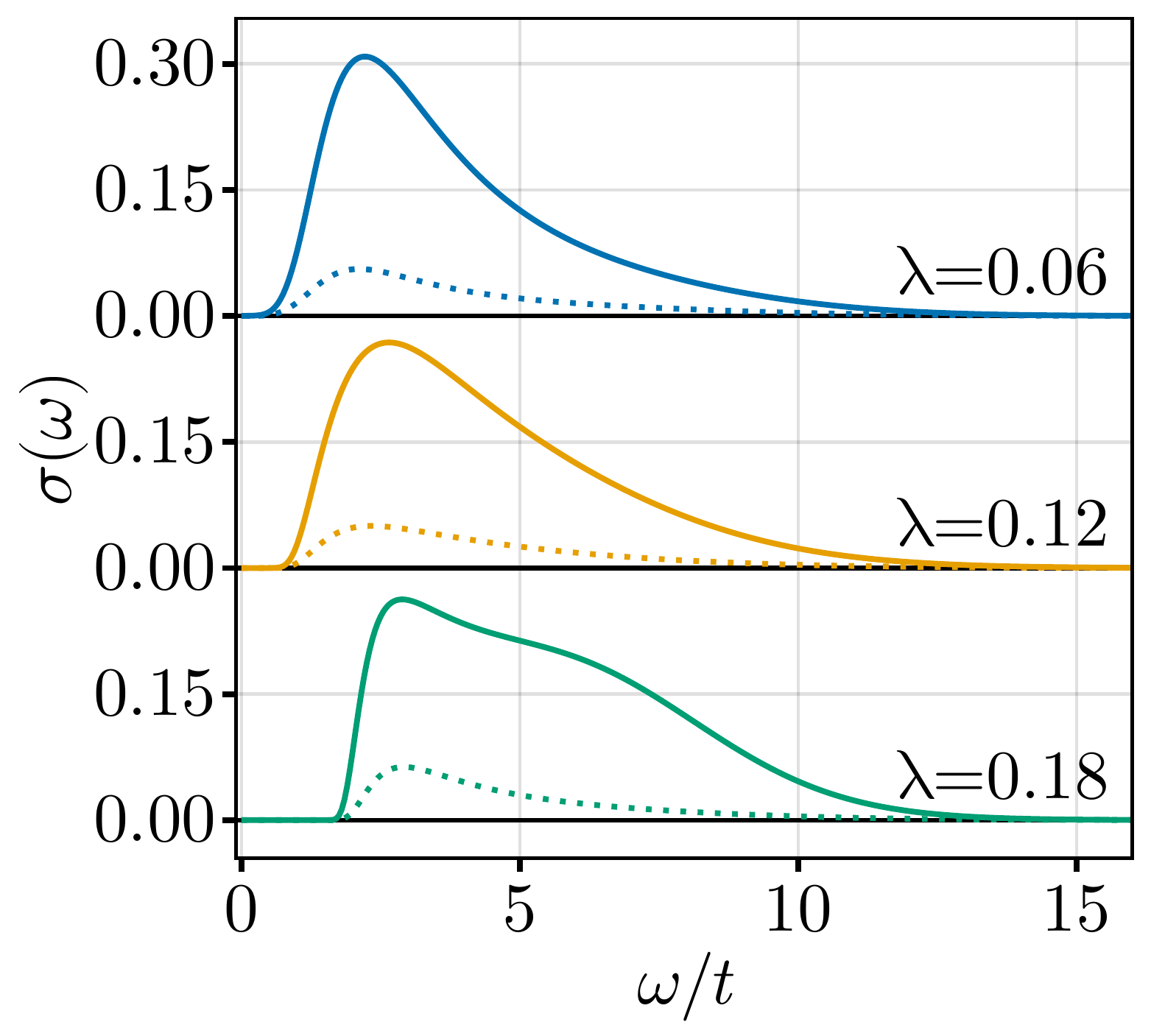}
    \caption{The optical conductivity for the \gls*{SSHH} model. We obtained these results using the same parameters as in Fig.~\ref{fig:Aw}. The solid and dotted lines correspond to the conductivity in the $\hat{x}$- and $\hat{y}$-directions, respectively.}
    \label{fig:cond}
\end{figure}

Next, we examine the dynamical properties of the strained model in the \gls*{AFM} and \gls*{BOW}, as well as the coexistence region. To this end, Fig.~\ref{fig:Aw} plots the single-particle spectral function along the high-symmetry cuts of the first Brillouin zone for fixed $\delta =0.3t$, $U=5t$, and $\Omega_0=0.5t$, obtained here on larger $12\times 12$ clusters. We vary the strength of the \gls*{eph} coupling to tune the system from the \gls*{AFM} phase [$\lambda = 0.06$, Fig.~\ref{fig:Aw}(a)] through the coexistence region [$\lambda = 0.12$, Fig.~\ref{fig:Aw}(b)] to the \gls*{BOW} phase [$\lambda = 0.18$, Fig.~\ref{fig:Aw}(c)]. In all three phases, we observe a clear gap in both the spectral function and the density of states. Fig.~\ref{fig:cond} plots the corresponding optical conductivities for these three cases, confirming that these gaps result from insulating behavior rather than bipolaron formation~\cite{Kovač2025, neuhaus2026}. 

The strain field clearly affects the dispersion of the electronic states in all three phases shown in Fig.~\ref{fig:Aw}. In particular, the states disperse more rapidly with energy along $\Gamma$-$X$ than along $\Gamma$-$Y$. This difference is already present in the \gls*{AFM} phase, where the lattice distortions associated with the \gls*{BOW} phase have not formed, and therefore reflect the large value of the applied strain $\delta = 0.3t$. We do not observe any significant changes of the dispersion in either the coexistence or \gls*{BOW} regions, which indicates that the additional changes in the effective hopping due to the lattice distortion are smaller than the changes introduced by the strain. This behavior is also reflected in the magnitude and anisotropy of the optical conductivity shown in Fig.~\ref{fig:cond}, where the magnitude of the conductivity at finite frequencies is four to five times larger along the $x$-direction compared to the $y$-direction. 

\section{Discussion and Conclusions}\label{sec:conclusions}

We have determined both the ground state phase diagram of the strained \gls*{SSHH} Hamiltonian, as well as the location of the finite temperature \gls*{BOW} transition at selected parameter values. Our phase diagram quantifies the effect of hopping anisotropy on the low temperature \gls*{AFM}-\gls*{BOW} competition.  The thin coexistence region which occurs in the isotropic case survives to non-zero $\delta=0.3t$.

We also explored the effect of anisotropy on the finite-temperature BOW transition.  Here we find $T_\mathrm{c}$ is almost constant for $0 < \delta < 0.6$, in sharp contrast to the Holstein case, where $T_\mathrm{c}$ decreases by at least a factor of four over that range~\cite{Cohen-Stead2019}. We attribute this to the presence of two competing effects:  The reduction in the possible ordered \gls*{BOW} phases from four to two drives an increase in $T_\mathrm{c}$ due to entropy reduction.  On the other hand, $\delta >0$ also leads to quasi-one-dimensional character, and this dimensional reduction lowers $T_\mathrm{c}$.  Since the anisotropy affects both the \gls*{eph} independent hopping in the $y$ direction as well as the \gls*{eph} contribution to the hopping the chains completely decouple at $\delta =1$.

A promising avenue for future work would be to examine the crossover from fourfold \gls*{BOW} criticality in the unstrained model to the ${\cal Z}_2$ Ising criticality induced by strain. If the unstrained transition is at the four-state-Potts point, the corresponding exponents are \(\gamma=7/6\) and \(\nu=2/3\), whereas the strained transition is expected to have the \gls*{2D} Ising exponents \(\gamma=7/4\) and \(\nu=1\). However, if the unstrained system lies on the Ashkin-Teller critical line, $\nu$ can take intermediate values while the ratio $\gamma/\nu=7/4$ remains fixed. This task is made more challenging because $\gamma/\nu$ is the same in all cases.  Thus, a full scaling collapse where the temperature is also scaled ($L^{1/\nu} (T-T_\mathrm{c})$) must be attempted, requiring an assessment of the quality of the overlap of noisy data with {\it two} adjustable parameters, $\nu$ and $T_\mathrm{c}$.

\section*{Acknowledgments}
The work of J.H.~and R.T.S.~was supported by the grant DOE DE-SC0014671 funded by 
the U.S. Department of Energy, Office of Science. Work at the University of 
Tennessee (code development, analytic continuation, writing) was supported by the 
National Science Foundation under grant No. OAC-2410280. 

\bibliography{StrainedSSHH}


\end{document}